\documentclass{article}[12pt]
\usepackage{amssymb,latexsym, amsmath,color,amsthm}
\usepackage{url, enumerate,comment,framed}
\usepackage[color=yellow]{todonotes}
\usepackage[margin=2cm]{geometry}
\newtheorem{remark}{Remark}
\newtheorem{example}{Example}
\def\MM#1{{\bf{#1}}}
\def\ep{{\epsilon }}
\def\bl{{\mathbf {l} }}
\def\bm{{\mathbf {m} }}

\def\bx{{\mathbf {x} }}

\def\bu{{\mathbf {u} }}
\def\bv{{\mathbf {v} }}

\def\bU{{\mathbf {U} }}

\def\bbR{{\mathbb {R} }}

\def\bsxi{{\boldsymbol {\xi} }}

\def\bszeta{{\boldsymbol {\zeta} }}

\def\dd{{\mathsf d}_t}
\def\p{{\partial }}
\def\qbar{{ \mathbf{\bar{q}} }}

\numberwithin{equation}{section}

\title{Stochastic Parametrization of the Richardson Triple}
\author{Darryl D. Holm \\ \normalsize
Mathematics Department, Imperial College, London, United Kingdom
} 
\date{
\\Hamilton--Pontryagin variational principle; 
\\stochastic fluid dynamics
\\ \bigskip
\begin{center} 
Big whorls have little whorls,
\\that feed on their velocity,
\\and little whorls have lesser whorls,
\\and so on to viscosity\bigskip
\\--- L. F. Richardson (1922)
\end{center}   
}

\begin{document}
\maketitle
\makeatother

\begin{abstract}
A Richardson triple is an ideal fluid flow map $g_{t/\ep,t,\ep t} = h_{t/\ep}k_t l_{\ep t}$ composed of three smooth maps with separated time scales: slow, intermediate and fast; corresponding to the big, little, and lesser whorls in Richardson's well-known metaphor for turbulence. Under homogenisation, as $\lim \ep\to0$, the composition $h_{t/\ep}k_t $ of the fast flow and the intermediate flow is known to be describable as a single stochastic flow $\dd g$. The interaction of the homogenised stochastic flow $\dd g$ with the slow flow of the big whorl is obtained by going into its non-inertial moving reference frame, via the composition of maps $(\dd g)l_{\ep t}$.  This procedure parameterises the interactions of the three flow components of the Richardson triple as a single stochastic fluid flow in a moving reference frame. 

The Kelvin circulation theorem for the stochastic dynamics of the Richardson triple reveals the interactions among its three components. Namely, (i) the velocity in the circulation integrand acquires is kinematically swept by the large scales; and (ii) the velocity of the material circulation loop acquires additional stochastic Lie transport by the small scales. The stochastic dynamics of the composite homogenised flow is derived from a stochastic Hamilton's principle, and then recast into Lie-Poisson bracket form with a stochastic Hamiltonian. Several examples are given, including fluid flow with stochastically advected quantities, and rigid body motion under gravity, i.e., the stochastic heavy top in a rotating frame.
\end{abstract}

\tableofcontents

\section{Introduction}

\subsection{The need for stochastic parametrization in numerical simulation}

\begin{figure}[h]
\centering
{\includegraphics[width=0.75\textwidth]{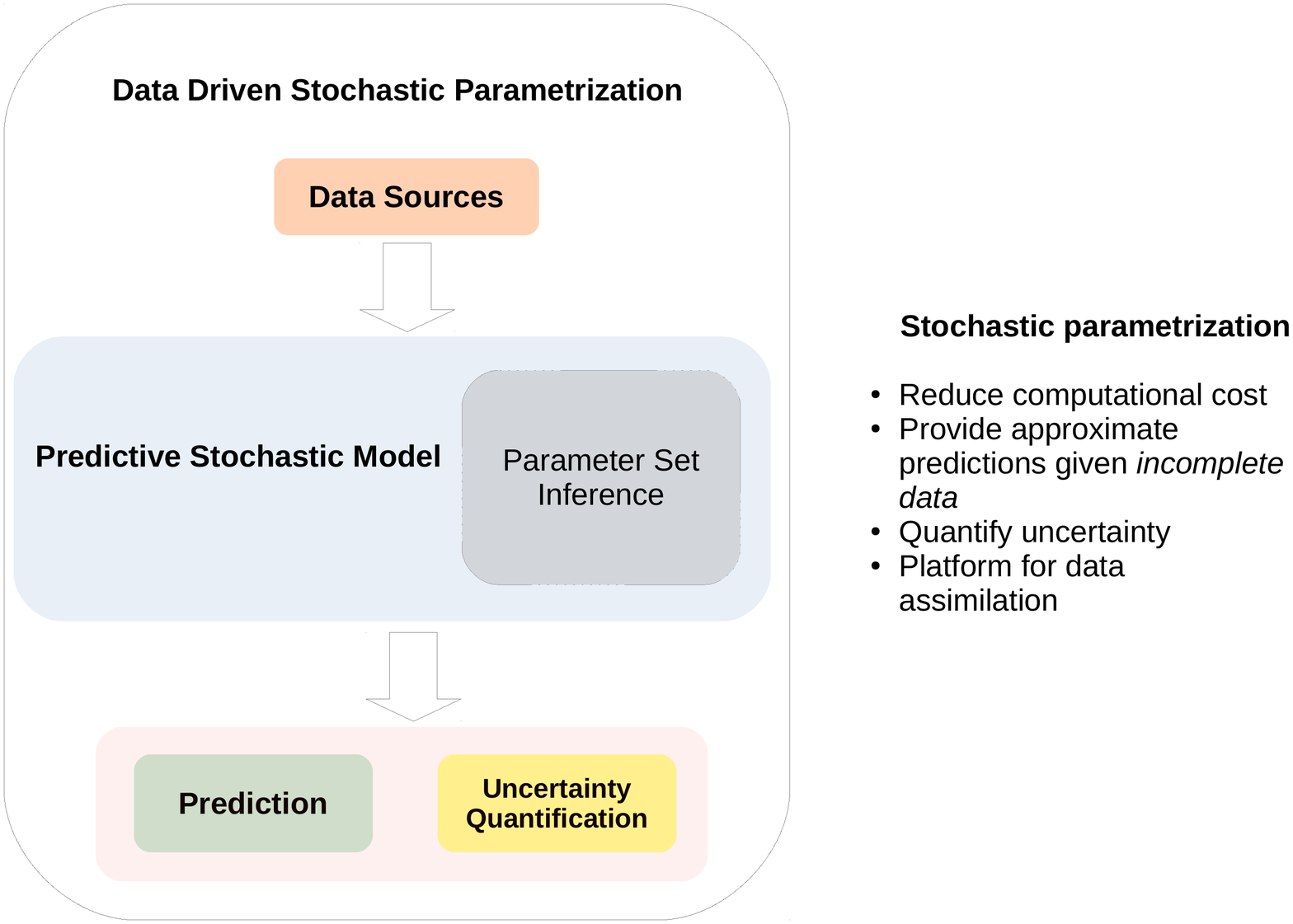} \label{fig:StochParamPic}%
}
\caption{A simplified flow chart of modelling by stochastic parametrisation.}%
\end{figure}

\paragraph{What is stochastic parametrization?}
Applications in the development of numerical simulation models for prediction of weather and climate have shown that averaging  the equations in time and/or filtering the equations in space in an effort to obtain a collective representation of their dynamics is often not enough to ensure reliable prediction. For better reliability, one must also estimate the uncertainty in the model predictions; for example, the variance around the space and time mean solution. This becomes a probabilistic question of how to model the sources of error which may be due to neglected and perhaps unknown or unresolvable effects in computational simulations. The natural statistical approach is to model these neglected effects as noise, or stochasticity. This in turn introduces the problem of \emph{stochastic parametrization}, \cite{Berner-etal2017,ChHa2009,PaSt2008,Palmer-etal2009}. Stochastic parametrization is the process of inference of parameters in a predictive stochastic model for a dynamical system, given partial observations of the system solution at a discrete sequence of times, see Figure \ref{fig:StochParamPic}. Stochastic parametrization is intended to: (i) reduce computational cost by constructing effective lower-dimensional models; (ii) enable approximate prediction when measurements or estimates of initial and boundary data are incomplete; (iii) quantify uncertainty, either in a given numerical model, or in accounting for unknown physical effects when a full model is not available; and (iv) provide a platform for data assimilation to reduce uncertainty in prediction via numerical simulation. Stochastic parametrization is often accomplished by introducing stochastic perturbations that represent unknown factors, preferably while still preserving the fundamental mathematical structure of the full model. For extensive recent reviews of stochastic parametrization for climate prediction, see, e.g., \cite{Berner-etal2017,Berner-etal2012}. For reviews of stochastic parametrization from the viewpoint of applied mathematics, see, e.g., \cite{Franzke-etal2015,MaFrKh2008}.

\subsection{Recent progress in the mathematics of stochastic parametrization}
\paragraph{Homogenization of multi-time processes leads to stochasticity.}
In recent developments aimed at including stochastic parametrization at a basic mathematical level, a variety of stochastic Eulerian fluid equations of interest in geophysical fluid dynamics (GFD) were derived in \cite{Holm2015} by applying Hamilton's variational principle for fluids \cite{HoMaRa1998}, and assuming a certain Stratonovich form of stochastic Lagrangian particle trajectories. The same stochastic dynamics of Lagrangian trajectories that had been proposed in \cite{Holm2015} was derived constructively in \cite{CoGoHo2017} by applying rigorous homogenization theory to a multi-scale, slow-fast decomposition of the deterministic Lagrangian flow map into a rapidly fluctuating, small-scale map acting upon a slower, larger-scale map for the mean motion. The derivation obtained using this slow-fast composition of flow maps in \cite{CoGoHo2017} of the same stochastic Lagrangian trajectories that had been the solution Ansatz for \cite{Holm2015} was achieved by applying multi-time homogenization theory to the vector field tangent to the flow of the slow-fast composition of flow maps; thereby obtaining stochastic particle dynamics for the resolved transport by the mean velocity vector field. 
In applying rigorous homogenization theory to derive the stochastic Lagrangian trajectories assumed in \cite{Holm2015}, the authors of \cite{CoGoHo2017} needed to assume mildly chaotic fast small-scale dynamics, as well as a \emph{centering condition} under which the mean displacement of the fluctuating deviations would be small, when pulled back to the mean flow \cite{PaSt2008}.  

\paragraph{Stochastic Advection by Lie Transport (SALT).}
The theorems for local-in-time existence and uniqueness, as well as a Beale-Kato-Majda regularity condition were proven in \cite{CrFlHo2017} for the stochastic Eulerian fluid equations for 3D incompressible flow which had been derived variationally in \cite{Holm2015} for stochastic Lagrangian trajectories. In addition, the stochastic Eulerian fluid equations for 3D incompressible flow derived in \cite{Holm2015} from a stochastically constrained variational principle were re-derived in \cite{CrFlHo2017} from Newton's force law of motion via Reynolds transport of momentum by the stochastic Lagrangian trajectories. Thus, the variational derivation in \cite{Holm2015} and the Newton's Law derivation for fluids in \cite{CrFlHo2017} was found to result in the same stochastic Eulerian fluid equations. Mathematically speaking, this class of fluid equations introduces Stochastic Advection by Lie Transport (SALT). SALT modifies the Kelvin circulation theorem so that its material circulation loop moves along the stochastic Lagrangian trajectories, while its integrand still has the same meaning (which is momentum per unit mass, a covariant vector) as for the deterministic case.  As for applications in stochastic parameterisation, the inference of model parameters for the class of SALT fluid models discussed here has recently been implemented for numerical simulations of Euler's equation for 2D incompressible fluid flows in \cite{Cotter-etal-2018a} and for simulations in 2D of 2-layer quasigeostrophic equations in \cite{Cotter-etal-2018b}.

\paragraph{Comparison of SALT with other approaches of stochastic parametrization.}
Of course, many other approaches to stochastic analysis and parameterisation of fluid equations have previously been investigated, particularly in the analysis of the stochastic Navier-Stokes equations; see. e.g., \cite{MiRo2004}.
Among the papers whose approaches are closest to the present work, we cite \cite{ArChCr2014} for the Euler-Poincar\'e variational approach, and \cite{Me2014,Re2017,ReMeCh2017I,ReMeCh2017II,ReMeCh2017III} for the method of location uncertainty (LU). In many ways, the LU approach is similar in concept to the SALT approach. In particular, the stochastic transport of scalars is identical in both LU and SALT models. However, the motion equation for evolution of momentum as well as the advection equations for non-scalar quantities in SALT models involve additional transport terms arising from transformation properties that are not accounted for in LU models. These additional terms for SALT models preserve the Kelvin circulation theorem, as well as the analytical properties of local existence, uniqueness and regularity shown in \cite{CrFlHo2017}. The additional stochastic transport terms for SALT models do not preserve energy, although they do preserve the Hamiltonian structure of ideal fluids. However, the explicit space and  time dependence in the stochastic Hamiltonian for fluid dynamics with SALT precludes the conservation of total energy and momentum. In contrast, the LU models do conserve total energy, although they do not preserve the Hamiltonian structure of ideal fluid dynamics. The LU models also do not possess a Kelvin circulation theorem, except in the case of incompressible flows in two dimensions, where the vorticity undergoes scalar advection. Energy conservation has also been a fundamental tenet of many other stochastic GFD models, such as stochastic climate models, \cite{MaTivdEi2003,MaFrKh2008,Franzke-etal2015,GoCrFr2016}. However, the SALT models show that stochastic parameterisation of unknown dynamical processes need not conserve energy, in general.

\subsection{Objectives of this work}
In the remainder of this paper we aim to extend the 2-level slow-fast stochastic model that was proposed in \cite{Holm2015}, derived in \cite{CoGoHo2017} and analysed in \cite{CrFlHo2017} to add another, slower, component of dynamics, inspired by LF Richardson's cascade metaphor. For this, we will use the transformation properties introduced in \cite{HoMaRa1998} and developed for the deterministic fluid dynamics of multiple spatial scales in \cite{HoTr2012}. The basic idea of the multi-space-scale approach of \cite{HoTr2012} is that each successively smaller scale of motion regards the previous larger scale as a Lagrangian reference frame, in which the mean of the smaller fluctuations vanishes. Here, we use Richardson's metaphor of ``whorls within whorls", interpreted in \cite{HoTr2012} as nested Lagrangian frames of motion, to extend the stochastic fluid model proposed in \cite{Holm2015} and derived in \cite{CoGoHo2017}  from 2 time-levels to 3 time-levels. 

The multi-space-scale approach of \cite{HoTr2012} follows one step in Richardson's cascade metaphor \cite{Ri1922}, which describes fluid dynamics as a sequence of whorls within whorls, in which each space/time scale is carried by a bigger/slower one. LF Richardson's cascade in scales (sizes) of nested interacting ``whorls" \dots ``and so on to viscosity", together with GI Taylor's hypothesis of  ``frozen-in turbulence" \cite{Ta1938} have inspired a rich variety of turbulence closure models based on multi-scale averaging for obtaining the dynamics of its collective quantities. 
The variety of multi-scale turbulence closure models include multifractal models \cite{Benzi-etal1984}, scale-similar wavelet models \cite{Argoul-etal1989}, self-similar moments of vorticity \cite{JDG2012}, shell models based on the renormalization group \cite{Eyink1993}, Large Eddy Simulation (LES) approaches \cite{MeKa2011}, Lagrangian Averaged Navier-Stokes (LANS) models \cite{Holm-etal-2005}, multi-scale turbulence models based on convected fluid microstructure \cite{HoTr2012} and many more.

To understand the implications of Richardson's cascade metaphor for stochastic parameterisation, one begins by understanding that each smaller/faster whorl in the sequence is not ``frozen'' into the motion of the bigger/slower one, as GI Taylor's hypothesis had once suggested \cite{Ta1938}.  Instead, each smaller/faster whorl also acts back on the previous bigger/slower one. Namely, while each successively smaller/faster scale of motion does regard the previous larger scale as a Lagrangian reference frame or coordinate system, it also acts back on the bigger/slower ones, as both a ``Reynolds stress'' and an enhanced transport of fluid. Reynolds stress is another classical idea from the history of turbulence modelling. Indeed, a multi-scale expression for Reynolds stress arose in \cite{HoTr2012} from a simple ``kinematic sweeping assumption'', that the fluctuations are swept by the large-scale motion and they have \emph{zero mean in the Lagrangian frame moving with the large-scale velocity}.  The latter zero mean assumption was also required, as the ``centering condition'', for the multi-time homogenisation treatment of stochastic parameterisation in \cite{CoGoHo2017}, which led to the stochastic solution Ansatz for the derivation of the SALT models introduced in \cite{Holm2015}. 

The stochastic advective transport terms arising in the SALT models of stochastic parameterisation comprise the sum of the divergence of a stochastic Reynolds stress and a stochastic line-element stretching rate. This additional stretching term in the SALT models is essential in preserving the Kelvin circulation theorem for the total solution. 

The extra stochastic line-element stretching arising from transport by stochastic Lagrangian trajectories in the SALT models represents their main difference from both the LU models and the class of turbulence models obtained by averaging the fluid equations in time and/or filtering the equations in an Eulerian sense in space in an effort to obtain a collective representation of their dynamics.
%
%
%
%
%

As we shall see, the line-element stretching term appearing in the SALT models arises from the transformation properties of the momentum density (a 1-form density) which transforms by pull-back under the smooth invertible map $g_t$ from the Lagrangian reference configuration of the fluid to its current Eulerian spatial configuration.
This transformation of the covariant vector basis for momentum density must also be included in deriving the fluid flow equations from Newton's law of force and motion. For example, see \cite{CrFlHo2017} for the derivation of the SALT Euler equations in \cite{Holm2015} from an application of the Reynolds transport theorem to Newton's law of force and motion. According to Newton's force law, the time derivative of the momentum in a volume of fluid equals the integrated force density applied to that volume of fluid, as it deforms, following the stochastic transport by the fluid flow.  This deformation again produces the SALT stretching term, which would be missed by simply applying space or time averaging to the Eulerian fluid equations, or by ignoring the transformation properties of the geometrical basis of momentum as a 1-form density under the deformation. 

\section{Background: Transformation properties of multi-time-scale motion}

In the previous section, we have explained that the line-element stretching term in the SALT models which ensures their representation as a stochastic Kelvin circulation theorem arises from the transformation properties of the momentum density (a 1-form density) under the smooth invertible maps (diffeomorphisms), regarded as a Lie group under composition of functions defined on an $n$-dimensional manifold $M$, the domain of flow. Here we discuss these transformations in detail, in preparation for using them throughout the remainder of the paper. For background about transformations by smooth invertible maps of tensor spaces defined on smooth manifolds, see, e.g., \cite{MaHu1983,HoScSt2009}.

Fluid momentum density $m\in \Lambda^1(M)\otimes \Lambda^n(M)$ on the manifold $M$ is a covariant vector density (that is, a 1-form density), which can be written in coordinates on $M$ as, e.g., $m=\bm(\bx,t)\cdot d\bx\otimes d^nx$.  1-form densities are dual in $L^2$ to vector fields $v\in\mathfrak{X}(M)$, interpreted physically as fluid velocities. The duality between 1-form densities and vector fields is represented by the $L^2$ pairing, 
\[
\langle m,v\rangle = \int  \bm\cdot \bv\,d^nx
\,.
\]
This $L^2$ duality is natural from the viewpoint of Hamilton's principle $\delta S = 0$ with $S=\int_a^b \ell(v)\,dt$ for ideal fluids with a Lagrangian $\ell: \mathfrak{X}\to \bbR$, since $\delta\ell(v) = \langle \delta \ell/\delta v\,,\, \delta v\rangle$ and the variational derivative of the Lagrangian with respect to the fluid velocity is identified with the fluid momentum density; that is, $\delta \ell/\delta v=:m$.

\paragraph{Pull-back and push-forward of a smooth flow acting on momentum and velocity.}
The time-dependent Lagrange-to-Euler map $g_t$ satisfying $g_sg_t=g_{s+t}$ represents fluid flow on the manifold $M$ as the composition of the flow of the diffeomorphism $g$ from time $0\to t$, composed with the flow from time $t\to t+s$. 

The momentum $m$ transforms under a diffeomorphism $g\in {\rm Diff}(M)$ by \emph{pull-back}, denoted 
\[
g_t^*m={\rm Ad}^*_{g_t}m,
\quad\hbox{that is,}\quad
g_t^*m= m\circ g_t\,,
\]
while a vector field $v$ undergoes the dual transformation; that is, by \emph{push-forward}, denoted 
\[
g_{t*}v={\rm Ad}_{g_t}v
\quad\hbox{that is,}\quad
g_{t*}v = Tg_t\circ  v\circ g_t^{-1}
\,,\quad\hbox{or, in components,}\quad
(g_{t*}v)^i(g_t\bx) = \frac{\p (g_t\bx)^i}{\p x^j}v^j(\bx)
\,.
\] 
The operations ${\rm Ad}_{g_t}$ (push-forward by $g_t$) and its dual ${\rm Ad}^*_{g_t}$ (pull-back by $g_t$) provide the \emph{adjoint and coadjoint representations} of ${\rm Diff}(M)$, respectively. Namely, for any $g,h\in {\rm Diff}(M)$, we have
\[
{\rm Ad}_g{\rm Ad}_h={\rm Ad}_{gh}
\quad\hbox{and}\quad
{\rm Ad}^*_{g^{-1}}{\rm Ad}^*_{h^{-1}}={\rm Ad}^*_{(gh)^{-1}} 
\,.
\]
The duality between ${\rm Ad}_{g_t}$ and ${\rm Ad}^*_{g_t}$ may be expressed in terms of the $L^2$ pairing $\langle \,\cdot\,,\,\cdot\,\rangle: \Lambda^1(M)\otimes \Lambda^n(M) \times \mathfrak{X}(M)\to\mathbb{R}$ as 
\begin{align}
\langle m, {\rm Ad}_{g_t}v\rangle = \langle {\rm Ad}^*_{g_t}m, v\rangle
\,.
\label{Ad-coAD}
\end{align}
This expression linearises at the identity, $t=0$, upon using the following formulas, for a fixed vector field, $\nu$, that
\begin{align}
\frac{d}{dt}\Big|_{t=0}{\rm Ad}_{g_t}\nu={\rm ad}_{\xi}\nu = {\cal L}_\xi \nu =-[\xi,\nu]
\,,\label{Ad-ad}
\end{align}
where $\xi=\dot{g}_tg_t^{-1} \in\mathfrak{X}(M)$,  and  $-[\xi,v] = \xi(v) - v(\xi)$ is the Lie bracket between vector fields. Likewise,  we have the linearisation,
\begin{align}
\frac{d}{dt}\Big|_{t=0}{\rm Ad}^*_{g_t}v={\rm ad}^*_{\xi}v= {\cal L}_\xi m\,,
\label{coAd-coad}
\end{align}
where ${\cal L}_\xi$ denotes the \emph{Lie derivative} with respect to the right-invariant vector field $\xi=\dot{g}_tg_t^{-1}$.
Together, equations \eqref{Ad-ad} and \eqref{coAd-coad} provide the following dual relations,
\begin{align}
\langle m, {\rm ad}_\xi v\rangle = \langle m, {\cal L}_\xi v\rangle 
= \langle {\cal L}^T_\xi m, v\rangle = \langle {\rm ad}^*_\xi m,  v\rangle
\,,
\label{ad-adstar-dual}
\end{align}
Relations \eqref{Ad-coAD} -- \eqref{ad-adstar-dual} facilitate the computations below in the addition of velocities in moving from one frame of motion to another, and they will be used several times later in the variational calculations in section \ref{Var-form-sec}. 

\paragraph{Flows of two time scales in slow/fast fluid systems.}
Consider the slow/fast flow given by 
\begin{align}
g_{t/\ep,t} = h_{t/\ep}k_t \,,
\label{2comp-flow}
\end{align}
obtained by composing a fast time flow $h_{t/\ep}$ parameterised by $t/\ep$ with $\ep\ll1$, with a slower time flow $k_t$ parameterised by time $t$. The Eulerian velocity for the composite flow \eqref{2comp-flow} is given by computing with the chain rule, in which over-dot denotes partial time derivative, e.g., $\dot{g}_{t/\ep}=\p_t g_{t/\ep}$,
\begin{align}
\begin{split}
\dot{g}_{t/\ep,t} g_{t/\ep,t}^{-1} 
&= \left(\dot{h}_{t/\ep}h_{t/\ep}^{-1} \right)
+ \frac{\p h_{t/\ep}}{\p k_t} (\dot{k}_t k_t^{-1})h_{t/\ep}^{-1}
\\&=:  \left(\dot{h}_{t/\ep}h_{t/\ep}^{-1} \right)
+ {\rm Ad}_{h_{t/\ep}}(\dot{k}_t k_t^{-1})
\,.
\end{split}
\label{chain-rule}
\end{align}
If $h_{t/\ep}$ is a near-identity transformation, so that $h_{t/\ep} = Id + \zeta_{t/\ep}$ as in \cite{CoGoHo2017}
and one defines the velocity $\dot{k}_t k_t^{-1} \bx = \bu(\qbar(\bl,t),t)$ along a Lagrangian trajectory of the slower time flow, $\bx=\qbar(\bl,t)=k_t\bl$, then, upon recalling that {\rm Ad} acts on vector fields by push-forward,  we have that
\begin{align}
\begin{split}
{\rm Ad}_{h_{t/\ep}}(\dot{k}_t k_t^{-1}) \bx
&= {\rm Ad}_{(Id + \zeta_{t/\ep})} \bu(\qbar(\bl,t) \,,\,t)
= {(Id + \zeta_{t/\ep})}_* \bu(\qbar(\bl,t),t) 
\\&= \bu\Big(\qbar(\bl,t) + \zeta_{t/\ep}\circ \qbar(\bl,t) \,,\, t\Big)
= \bu \big( \bx + \bszeta( \bx,t/\ep)\,,\,t\big)
\,.
\end{split}
\label{vel-xform}
\end{align}
Now, the total velocity of the composite flow is the time derivative of the sum of two flows, since  $g_{t/\ep,t} = (Id + \zeta_{t/\ep})k_t = k_t + \zeta_{t/\ep}k_t$. Consequently, by the chain rule, the total velocity is given by
\begin{align}
\begin{split}
\dot{g}_{t/\ep,t} g_{t/\ep,t}^{-1} \bx
&= \frac{\p}{\p t}\Big|_\bl \Big(\qbar(\bl,t) + \bszeta\big( \qbar(\bl,t),t/\ep\big)\Big)
\\&= \mathbf{\dot{\qbar}}(\bl,t) + \left(\mathbf{\dot{\qbar}}\cdot \frac{\p}{\p \qbar}\right)\bszeta(\qbar(\bl,t),t/\ep) 
+ \mathbf{\dot{\bszeta}} (\qbar(\bl,t),t/\ep)
\\& = \bu(\bx,t) + (\p_t + \bu\cdot\nabla ) \bszeta(\bx,t/\ep)
=:  \bu(\bx,t) + \frac{d}{dt}\Big|_\bu \!\!\bszeta(\bx,t/\ep)
\,.
\end{split}
\label{2time-vel}
\end{align}
According to the centering condition mentioned earlier, the mean displacement of the fluctuating deviations vanishes, when pulled back to the mean flow. Upon denoting the mean as an over-bar $\overline{(\,\cdot\,)}$, the centering condition of \cite{CoGoHo2017} may be expressed in the current notation as 
\begin{align}
\overline{g}_{t/\ep,t} = k_t + \overline{\zeta_{t/\ep}k_t} = k_t
\,.
\label{centr-cond}
\end{align}
In particular, this implies that the mean velocity is given by
\begin{align}
\dot{\overline{g}}_{t/\ep,t} \overline{g}_{t/\ep,t}^{\,-1} \,\bx = \dot{k}_t k_t^{-1}\bx = \bu(\bx,t) = \mathbf{\dot{\qbar}}(\bl,t) 
\,.
\label{mean-vel}
\end{align}
Consequently, we may interpret the velocity $\dot{k}_t k_t^{-1} \bx = \bu(\bx,t) = \bu(\qbar(\bl,t),t)$ as the velocity along the mean Lagrangian trajectory, which is defined as $\bx=\qbar(\bl,t)=k_t\bl$. Thus, the mean that has been defined by imposing the centering condition is the average over the fast time $t/\ep$, at fixed Lagrangian label, $\bl$. This is \emph{Lagrangian averaging} of the fluid transport velocity. For discussions of Lagrangian averaged fluid equations, see, e.g., \cite{AnMc1978,GiVa2018}. For Lagrangian averaged Hamilton's principles for fluids, see, e.g., \cite{GjHo1996}, and for an application of the latter to turbulence modelling, see \cite{Chen-etal1999}.

\paragraph{Homogenization of slow/fast fluid systems.}
In \cite{CoGoHo2017}, homogenization theory was employed to show that the expression for total velocity in \eqref{2time-vel} converges for $\ep\to 0$ to produce a certain type of stochastic Lagrangian dynamics on long time scales, provided the centering condition \eqref{centr-cond} holds. In particular, homogenization theory for deterministic multi-scale systems as developed in \cite{MelbourneStuart11,GottwaldMelbourne13,KellyMelbourne17} assures that under certain general conditions the slow $t$-dynamics of this deterministic multi-scale Lagrangian particle dynamics is described over long time-scales of $O(1/\ep^2)$ by the \emph{stochastic differential equation} \cite{CoGoHo2017},
\begin{equation}
  \label{eq:dq}
  \dd \MM{q} = \MM{u}(\MM{q},t)\,{d t} + \sum_{i=1}^m \bsxi_i(\MM{q})\circ {d W^i_t}\,.
\end{equation}
Here $\dd$ is the notation for a stochastic process, the $W_t^i$ are independent stochastic Brownian paths, and the $\bsxi_i(\MM{q})$ are time-independent vector fields, which are meant to be determined from data. For examples of the determination of the $\bsxi_i(\MM{q})$ at coarse resolution from fine resolution computational data, see \cite{Cotter-etal-2018a,Cotter-etal-2018b}.The symbol $(\,\circ\,)$ in the stochastic process in \eqref{eq:dq} denotes a cylindrical Stratonovich process.%
\footnote{Cylindrical Stratonovich processes and their properties are defined and discussed in \cite{Sc1988}.} Formula \eqref{eq:dq}  was the solution Ansatz in the variational principles for stochastic fluid dynamics used in the original derivation of the SALT models in \cite{Holm2015}. Formula \eqref{eq:dq} had appeared already in a variety of other studies of stochastic fluid dynamics and turbulence, as well. See, e.g., \cite{Holm2015,MiRo2004} and references therein.

\paragraph{Richardson triples: Flows with three time scales -- slow/intermediate/fast fluid systems.}
Next, we extend the temporal dynamics of the 2-component slow/fast flow in \eqref{2compflow} to introduce time dependence into the stationary Lagrangian reference frame for the slow/fast flow; so that the previous reference coordinate $\bl$ will acquire a  slow time dependence as, say, $\bl(\bx_0,\ep t) = l_t\bx_0$, for a reference fluid configuration with labels $\bx_0$. For the 3-component flow obtained fro  the composition of flows for fast time $t/\ep$, intermediate time $t$, and slow time $\ep t$ is given by, 
\begin{align}
g_{t/\ep,t,\ep t} = h_{t/\ep}k_t l_{\ep t}\,.
\label{3compflow}
\end{align}
In the 3-component flow, the previous Eulerian velocity addition formula for a 2-component flow \eqref{2time-vel} now acquires an additional velocity in the summand, whose dependence on slow time, $\dot{l}_{\ep t} l_{\ep t}^{-1}\bx= \bU(\bx, \ep t)$ must be determined, or specified, so that, 
\begin{align}
\begin{split}
\dot{g}_{t/\ep,t,\ep t}g_{t/\ep,t,\ep t}^{-1}\bx
&= 
{\rm Ad}_{h_{t/\ep}}{\rm Ad}_{k_t}(\dot{l}_{\ep t} l_{\ep t}^{-1}) \bx
= \bU \Big(\qbar(\bl(\bx_0,\ep t),t) + \bszeta\big( \qbar(\bl(\bx_0,\ep t),t),t/\ep\big)\, ,\ep t\Big)
\\&=
 \frac{\p}{\p t}\Big|_\bl \Big(\qbar(\bl(\bx_0,\ep t),t) + \bszeta\big( \qbar(\bl(\bx_0,\ep t),t),t/\ep\big)\Big)
\\&= \mathbf{\dot{\qbar}}(\bl(\bx_0,\ep t),t) + \left(\qbar\cdot \frac{\p}{\p \qbar}\right)\bszeta(\qbar(\bl(\bx_0,\ep t),t),t/\ep) 
+ \frac{\p\bszeta}{\p (t/\ep)}\ep^{-1}
\,.
\end{split}
\label{3comp-vel}
\,.\end{align}
Rather than expanding out the slow-time derivatives in $\ep t$ here, we will find it convenient to first take the limit $\ep\to0$ to recover the SALT equations and then transform Hamilton's principle into the prescribed moving reference frame of the slow flow, in order to compute the dynamics in the three separated time levels as stochastic dynamics at intermediate time in a slowly varying, moving reference frame, with velocity $\dot{l}_{\ep t} l_{\ep t}^{-1}\bx= \bU(\bx, \ep t)$. 

\paragraph{Interim summary.}
So far, we have begun by simplifying LF Richardson's cascade involving an infinite sequence of triples of big whorls, little whorls and lesser whorls, into a single triple of whorls, by composing a slow flow, an intermediate flow and a fast flow, as $g_{t/\ep,t,\ep t} = h_{t/\ep}k_t l_{\ep t}$ in equation \eqref{3compflow}. We then invoked the results of \cite{CoGoHo2017} that the homogenisation method as $\lim \ep\to0$ consolidates the composition $h_{t/\ep}k_t $ of the fast flow and the intermediate flow into a single stochastic flow $\dd g$. In the next section, we will model the interaction of the consolidated stochastic flow with the slow flow of the big whorl, by going into the big whorl's moving reference frame via the composition $(\dd g)l_{\ep t}$. 

In the remainder of the paper, (i) we will consider only a single Richardson triple of whorls and (ii) we will model the interactions of the components of the Richardson triple as a single stochastic fluid flow in a moving reference frame. If we were transforming the equations of motion into the moving reference frame directly from \eqref{3comp-vel}, the process might seem tedious. However, because we have Hamilton's principle available for the entire flow, the transformation will be straight forward and the resulting dynamics will reveal the interactions we seek among the three components of the Richardson triple, even when the big whorl is taken to be moving stochastically. 

\paragraph{The primary aim of the rest of the paper.}
We are interested in rephrasing the dynamics of the Richardson triple in terms of Kelvin's circulation theorem, both for interpretation of the result and in anticipation of eventually using stochastic parameterisation to model the effects of the larger and smaller whorls on the circulation of an intermediate size whorl, as spatially dependent stochastic processes. The effects of the smaller whorl are to be modelled using Stratonovich stochastic Lie transport as in \cite{Holm2015,CrFlHo2017}, while the effects of the larger whorl in the Richardson triple  are to be modelled using an Ornstein-Uhlenbeck (OU) process that boosts the total fluid motion into a non-inertial frame. Although we will work with the Ornstein-Uhlenbeck (OU) process, the velocity $\dot{l}_{\ep t} l_{\ep t}^{-1}\bx= \bU(\bx, \ep t)$ of the slowly varying reference frame could be represented by any other semimartingale, or even be deterministic, since it must be prescribed from outside information, to complete the dynamical description.

The present approach results in a stochastic partial differential equations (SPDE) for the motion of an intermediate flow, stochastically transported by the fast flow of the small whorl, in the non-inertial frame of the slowly changing big whorl. One may imagine that the big whorl represents synoptic, long-term dynamics, the intermediate whorl represents the dynamics of daily, or hourly interest, and the stochastic terms represent unresolvably fast dynamics whose spatial correlations have been determined and are represented by the functions $\bsxi_i(\bx)$.

This approach is intended to be useful in multi-scale geophysical fluid dynamics. For example, it may be useful in quantifying predictability and variability of sub-mesoscale ocean flow dynamics in a stochasticly parameterised frame of motion representing one, or several, unspecified mesoscale eddies interacting with each other. One could also imagine using this Richardson triple approach to assess the influence on synoptic large scale weather patterns on stochastic hurricane tracks over a few days. Although we have no examples to cite yet, we hope this approach may also be useful in investigating industrial flows, e.g., to quantify  uncertainty and variability in multi-scale flow processes employed in industry.

{\bf Plan of the paper.}

In section \ref{Var-form-sec}, we present a variational formulation of the stochastic Richardson triple. 
In the corresponding equations of fluid motion, two types of stochasticity appear. Namely, 
\begin{enumerate}[(i)]
\item
\emph{Non-inertial stochasticity} due to random sweeping by the larger whorls, which adds to the \emph{momentum density} as an OU process representing the slowly changing reference-frame velocity, and  
\item
\emph{Stochastic Lie transport} due to the smaller whorls, which is added to the \emph{transport velocity} of the intermediate scales as a cylindrical Stratonovich process.
\end{enumerate}

Example \ref{stoch-vort-eqn} in section \ref{Var-form-sec} treats the application of this framework to derive the corresponding stochastic  vorticity equation for the incompressible flow of an Euler fluid in 3D.  

In section \ref{Ham-form-sec}, we present the Lie--Poisson Hamiltonian formulation of the stochastic Richardson triple  derived in the earlier sections. We find that the Hamiltonian obtained from the Legendre transform of the Lagrangian in the stochastic Hamilton's principle in section \ref{Var-form-sec} also becomes stochastic. Nonetheless, the semidirect-product Lie--Poisson Hamiltonian structure of the deterministic ideal fluid equations \cite{HoMaRa1998} persists. 

This persistence of Hamiltonian structure implies the preservation of the standard potential vorticity (PV) invariants even for the stochastic Lie transport dynamics of fluids in a randomly moving reference frame. To pursue the geometric mechanics framework further, we show in section \ref{Ham-form-sec} that the Lie--Poisson Hamiltonian structure of ideal fluid mechanics is preserved by the introduction of either, or both types of stochasticity treated here. As a final example, we consider the application of these ideas to the finite dimensional example of the stochastic heavy top. This is an apt example, because of the well-known gyroscopic analogue with stratified fluids, as discussed, e.g., in \cite{Holm1986,Do2013} and references therein. 
Moreover, the effects of transport stochasticity alone on the dynamical behaviour of the heavy top in the absence of the OU noise, was investigated in \cite{ArCaHo2017}. 

\paragraph{Previous variational approaches to stochastic fluid dynamics.} A variational formulation of the stochastic Kelvin Theorem (based on the back-to-labels map) led to the derivation of the Navier-Stokes equations in \cite{CoIy2008}. The same approach was applied in \cite{Eyink2010}, to show that this Kelvin theorem, regarded as a stochastic conservation law, arises via Noether's theorem  from particle-relabelling symmetry of the corresponding action principle, when expressed in terms of Eulerian variables. The latter result was to be expected, because the classical Kelvin theorem is a universal expression which follows from particle-relabelling symmetry for the Eulerian representation of fluid dynamics. See, e.g., \cite{HoMaRa1998} for the deterministic case, where this result is called the Kelvin-Noether theorem.  

In citing these precedents, we note that the goal of the present work is to use the metaphor of the Richardson triple to inspire the derivation of SPDEs for stochastic fluid dynamics, as first done in \cite{Holm2015}. In contrast to  \cite{CoIy2008,Eyink2010}, it is not our intention to derive the Navier--Stokes equations in the present context.  For a discussion of the history of variational derivations of stochastic fluid equations and their relation to the Navier--Stokes equations, one should consult the original sources, some of which are cited in \cite{Holm2015}.

\section{Stochastic variational formulation of the Richardson triple}\label{Var-form-sec}

We will be considering two types of noise for modelling the Richardson triple as a single stochastic PDE. The first type of noise represents the large scale, possibly stochastic, sweeping by the larger whorl. The second one represents flow with stochastic transport by the smaller whorl. These two types of noise can both be implemented and generalized at the same time by invoking the reduced Hamilton--Pontryagin variational principle for continuum motion $\delta S = 0$ \cite{HoMaRa1998,Holm2011,HoScSt2009}, and choosing the following stochastic action integral \cite{Holm2015,GBHo2017},
\begin{equation}
S = \int \ell(u,a_0g^{-1},D)\,dt + \langle\,DR(x,t)\,,\,u\,\rangle dt
+
\langle\,\mu\,,\,\dd g\,g^{-1} - u\,dt - \xi(x)\circ dW_t\,\rangle\,.
\label{ActionInt-1}
\end{equation}
Here,  $g(t)\in {\rm Diff}(\mathbb{R}^3)$ is a stochastic time-dependent curve in the manifold of diffeomorphisms, ${\rm Diff}(\mathbb{R}^3)$; the quantities  $u(x,t)$, $\xi(x)$ and  $\dd g\,g^{-1}$ are Eulerian vector fields defined appropriately over the domain of flow, and $a_0\in V$ for a vector space $V$. The Lagrangian $\ell(u,a_0g^{-1},D)$ is a general functional of its arguments.

The co-vector Lagrange multiplier $\mu$ enforces the vector field constraint
\begin{equation}
\dd g \,g^{-1} = u(x,t)\,dt + \sum_i\xi_i(x)\circ dW^i_t
\,,\quad\hbox{with}\quad
x(x_0,t) = g(t)x_0
\,,
\label{VF-1}
\end{equation}
applied to the variations of the action integral in \eqref{ActionInt-1}, in which the brackets $\langle \,\cdot\,,\,\cdot\,\rangle$ denote $L^2$ pairing of spatial functions. The space and time dependence of the stochastic parameterisation in the constrained Lagrangian for the action integral \eqref{ActionInt-1} introduces explicit space and time dependence. Consequently, the stochastic parameterisation $\sum_i\xi_i(x)\circ dW^i_t$ in the flow map $g(t)$ and the reference frame velocity $R(x,t)$ prevents conservation of total energy and momentum. In addition, the presence of the initial condition $a_0$ for the Eulerian advected quantity $a(t)=a_0g^{-1}(t)$ restricts particle-relabelling symmetry to the isotropy subgroup of the diffeomorphisms that leaves the initial condition $a_0$ invariant.  In the deterministic case, this breaking of a symmetry to an isotropy subgroup of a physical order parameter leads to semidirect-product Lie--Poisson Hamiltonian structure. As we shall see, this semidirect-product structure on the Hamiltonian side persists for the type of stochastic deformations we use in the present  case, as well.  The density $D=D_0g^{-1}$ (a volume form) is also an advected quantity. However, we have written $D$ separately in the action integral \eqref{ActionInt-1}, because it multiplies the reference frame velocity, $R(x,t)$, which plays a special role in the total momentum.

\paragraph{Reference frame velocity $R(x,t)$.}
The term in the Lagrangian \eqref{ActionInt-1} involving $R(x,t)$ transforms the motion into a moving frame with co-vector velocity $R(x,t)$. The moving frame velocity $R(x,t)$ must be specified as part of the formulation of the problem to be investigated. Having this freedom, we choose to specify it here as a stochastic process, because doing so gives us an opportunity to show that inclusion of non-inertial stochastic forces is compatible with stochastic transport, which is the basis of the SALT approach. In taking this opportunity, we will use an Ornstein-Uhlenbeck (OU) process, although any other semimartingale process could also have been used. Namely, we will choose the frame velocity to be 
\begin{equation}
R(x,t) = \eta(x)N(t)  
\,,
\label{refvel-def}
\end{equation}
where $\eta(x)$ is a smooth spatially dependent co-vector (coordinate index down) and $N(t)$ is the solution path of the stationary Gaussian-Markov OU process whose evolution is given by the following stochastic differential equation,
\begin{equation}
\dd R(x,t) = \eta(x) \dd N(t)  
\,,\quad\hbox{with}\quad
\dd N = \theta (\overline{N} - N(t))\,dt + \sigma dW_t
\,,
\label{OU-1}
\end{equation}
with long-term mean $\overline{N}$, and real-valued constants $\theta$ and $\sigma$. The solution of \eqref{OU-1} is known to be
\begin{equation}
N(t) = e^{-\theta t}N(0) + (1 - e^{-\theta t})\overline{N}
+ \sigma \int_0^t e^{-\theta (t-s)}dW_s
\,,
\label{OUsoln-1}
\end{equation}
in which one assumes an initially normal distribution, 
$N(0)\approx {\cal N}(\overline{N},\sigma^2/(2\theta))$, with mean $\overline{N}$ and variance $\sigma^2/(2\theta)$. 

As a mnemonic, we may write 
\[
{\rm curl}\,R(x,t)=:2\Omega(x)N(t)
\,,
\]
to remind ourselves that $R(x,t)$ represents the velocity of a stochastically moving reference frame, and that ${\rm curl}\,\eta(x)=:2\Omega(x)$ suggests the Coriolis parameter for a spatially dependent angular rotation rate. That is, $u(x,t)$ is the fluid velocity of interest relative to a moving reference frame with stochastic velocity $R(x,t)$.

\paragraph{Stationary variations of the action integral.}
Taking stationary variations of the action integral in equation \eqref{ActionInt-1} yields, 
\begin{align}
\begin{split}
0 = \delta S = \int
\left\langle
\frac{\delta \ell}{\delta u} + DR(x,t) - \mu\,,\, \delta u
\right\rangle\,dt
&+
\Big\langle\,\delta\mu\,,\,\dd gg^{-1} - u\,dt - \xi(x)\circ dW_t\,\Big\rangle
\\&+
\Big\langle\,\mu\,,\,\delta\left(\dd gg^{-1}\right) \,\Big\rangle
+
\left\langle\,\frac{\delta \ell}{\delta a}\,,\,\delta a 
\,\right\rangle
\,dt
+
\left\langle\,\frac{\delta \ell}{\delta D}
+ \big\langle R\,,\,u\big\rangle \,,\,\delta D 
\,\right\rangle
\,dt
\,,
\end{split}
\label{Var-1}
\end{align}
with advected quantities $a=a_0g^{-1}$, whose temporal and variational derivatives satisfy similar equations; namely,
\begin{align}
\begin{split}
\dd a &= - a_0g^{-1}\dd gg^{-1}  
= - a\left(\dd gg^{-1} \right) 
= - \pounds_{\dd gg^{-1} }a
\,,\\
\quad\hbox{and}\quad
\delta a &= - a_0g^{-1}\delta g g^{-1}  
= - \pounds_{\delta gg^{-1} }a
:= - \pounds_{ w}a
\,,\quad\hbox{with}\quad
w:= \delta gg^{-1} 
\,.
\end{split}
\label{a-dyn-var-1}
\end{align}
As before, $\pounds_{(\cdot)}$ denotes Lie derivative with respect to a vector field. In particular, the density variation is given by $\delta D = -\,{\rm div} (D w)$.
A quick calculation of the same type yields two more variational equations,
\begin{align}
\begin{split}
\delta \left(\dd g g^{-1} \right) 
&= (\delta \dd g) g^{-1} - \delta gg^{-1} \dd gg^{-1} 
= (\delta \dd g) g^{-1} - w\dd gg^{-1} 
\,,\\
\dd \left(\delta g g^{-1}\right)  
&
= (\dd \delta g) g^{-1} - \dd gg^{-1} \delta g g^{-1}
= (\dd \delta g) g^{-1} - \dd gg^{-1} w
\,.
\end{split}
\label{g-var-1}
\end{align}
Taking the difference of the two equations in \eqref{g-var-1} yields the required expression for the variation of the vector field $\delta (\dd g g^{-1} ) $ in terms of the vector field, $w:= \delta gg^{-1}$; namely,
\begin{equation}
\delta \left(\dd g g^{-1} \right) 
=
\dd w - \left[\dd g g^{-1} ,w \right]
=
\dd w - {\rm ad}_{\dd g g^{-1}}w
\,,
\label{ad-1}
\end{equation}
where ${\rm ad}_{v}w:=vw-wv$ denotes the adjoint action (commutator) of vector fields. 
Finally, one defines the diamond $(\diamond)$ operation as \cite{HoMaRa1998}
\begin{equation}
\left\langle\,\frac{\delta \ell}{\delta a}\diamond a\,,\,w \right\rangle_\mathfrak{g}
:=
\left\langle\,\frac{\delta \ell}{\delta a}\,,\,- \pounds_{ w}a\,\right\rangle_V
\,,\label{diamond-def}
\end{equation}
in which, for clarity in the definition of diamond $(\diamond)$ in \eqref{diamond-def} we introduce subscripts on the brackets $\langle \,\cdot\,,\,\cdot\,\rangle_\mathfrak{g}$ and $\langle \,\cdot\,,\,\cdot\,\rangle_V$ to identify the symmetric, non-degenerate pairings. These pairings are defined separately on the Lie algebra of vector fields $\mathfrak{g}$ and on the tensor space of advected quantities $V$, with elements of their respective dual spaces, $\mathfrak{g}^*$ and $V^*$. 
Inserting these definitions into the variation of the action integral in \eqref{Var-1} yields (once again suppressing subscripts on the brackets)
\begin{align}
\begin{split}
0 = \delta S = \int
&\left\langle
\frac{\delta \ell}{\delta u} + DR(x,t) - \mu\,,\, \delta u
\right\rangle\,dt
+
\left\langle\,\delta\mu\,,\,\dd gg^{-1} - u\,dt - \xi(x)\circ dW_t\,\right\rangle
\\&+
\left\langle\,\mu\,,\,\left(\dd w - {\rm ad}_{\dd g g^{-1}}w\right) \,\right\rangle
+
\left\langle\,\frac{\delta \ell}{\delta a}\,,\,(- \pounds_{ w}a)\,\right\rangle\,dt
+
\left\langle\,
\frac{\delta \ell}{\delta D}
+ \big\langle R \,,\,u\big\rangle 
 \,,\,-\,{\rm div} (D w)
\,\right\rangle\,dt
\,.
\end{split}
\label{Var-2}
\end{align}
Integrating by parts in both space and time in \eqref{Var-2} produces
\begin{align}
\begin{split}
0 = \delta S &= \int
\left\langle
\frac{\delta \ell}{\delta u} + DR(x,t) - \mu\,,\, \delta u
\right\rangle\,dt
+
\Big\langle\,\delta\mu\,,\,\dd gg^{-1} - u\,dt - \xi(x)\circ dW_t\,\Big\rangle
\\&-
\Big\langle\,\dd\mu + {\rm ad}^*_{\dd g g^{-1}}\mu\,,\, w \,\Big\rangle
+
\left\langle\,\frac{\delta \ell}{\delta a}\diamond a\,,\,w
\,\right\rangle\,dt
+
\left\langle\,
D\,\nabla\left(\frac{\delta \ell}{\delta D}
+ \big\langle R \,,\,u\big\rangle \right)  \,,\,w
\,\right\rangle\,dt
+
\int\dd \langle\,\mu\,,\, w \,\rangle
\,,
\end{split}
\label{Var-s}
\end{align}
in which the last term vanishes at the endpoints in time and boundary conditions are taken to be homogeneous.

Putting all of these calculations together yields the following results for the variations in Hamilton's principle \eqref{Var-1}, upon setting the variational vector field, $w:= \delta gg^{-1}$ equal to zero at the endpoints in time, 
\begin{align}
\begin{split}
\delta u:\quad&
\frac{\delta \ell}{\delta u} + DR(x,t) = \mu   \,,
\\
\delta\mu:\quad&
\dd gg^{-1} = u\,dt + \xi(x)\circ dW_t
\,,
\\
\delta g:\quad&
\dd \mu + {\rm ad}^*_{\dd g g^{-1}}\mu 
= \frac{\delta \ell}{\delta a}\diamond a\,dt
+ D\,\nabla\left(\frac{\delta \ell}{\delta D}
+ \big\langle R \,,\,u\big\rangle \right) dt
\,,
\end{split}
\label{results-var-2}
\end{align}
where the advected quantities $a$ and $D$ satisfy 
\begin{equation}
\dd a = - \,\pounds_{\dd gg^{-1} }a
\,,\qquad
\dd D = - \,{\rm div}\,(D \, \dd g g^{-1}) = - \,\pounds_{\dd gg^{-1} }D
\,,
\label{advect-2}
\end{equation}
as in \eqref{a-dyn-var-1}. In the last line of \eqref{results-var-2} one defines the coadjoint action of a vector field $\dd g g^{-1}$ on an element $\mu$ in its $L^2$ dual space of 1-form densities as, cf. \eqref{ad-adstar-dual},
\begin{equation}
\left\langle\,{\rm ad}^*_{\dd g g^{-1}}\mu\,,\, w \,\right\rangle 
= 
\left\langle\,\mu\,,\,{\rm ad}_{\dd g g^{-1}}w \,\right\rangle
.
\label{co-ad-1}
\end{equation}
Equations \eqref{results-var-2} now have stochasticity in both the momentum density $\mu$ and in the Lie transport velocity $\dd g g^{-1}$, while the advection equation \eqref{advect-2} naturally still only has stochastic transport.
 
Conveniently, the coadjoint action ${\rm ad}_v^*$ of a vector field $v$ on a 1-form density is equal to the Lie derivative $\pounds_v$ of the 1-form density with respect to that vector field, so that%
\footnote{This coincidence explains why Lie derivatives appear in fluid motion equations.}
\begin{equation}
{\rm ad}^*_{\dd g g^{-1}}\mu = \pounds_{\dd g g^{-1}}\mu
\,.
\label{ad*-1}
\end{equation}
The advection of the mass density $D$ is also a Lie derivative, cf. \eqref{advect-2}. Consequently, the last line of \eqref{results-var-2} implies the following stochastic equation of motion in 3D coordinates,
\begin{equation}
\begin{split}
\dd (\mu/D) + \pounds_{\dd g g^{-1}}(\mu/D)
&=
\Big[\dd \mathbf{v} - \dd g g^{-1} \times {\rm curl}\,\mathbf{v}
+ \nabla \big(\dd g g^{-1} \cdot \mathbf{v}\big)  \Big]\cdot d\mathbf{x}  
\\&= 
\frac1D \frac{\delta \ell}{\delta a}\diamond a\,dt
+ d\left(\frac{\delta \ell}{\delta D}
+ \big\langle R \,,\,u\big\rangle \right) dt
\,,
\end{split}
\label{mu/D-eqn}
\end{equation}
where  we have defined the momentum 1-form (a co-vector) as
\begin{equation}
\mu/D := \frac1D
\left(\frac{\delta \ell}{\delta \mathbf{u}} + D\mathbf{R}(\mathbf{x},t) \right)
\cdot d\mathbf{x} =:\mathbf{v}\cdot d\mathbf{x}\,,
\label{v-def}
\end{equation}
and $\dd g g^{-1}$ is given by the variational constraint in equation \eqref{results-var-2}. 

\paragraph{Kelvin's circulation theorem.}
Taking the loop integral of equation \eqref{mu/D-eqn} allows us to write the motion equation compactly, in the form of Kelvin's circulation theorem,
\begin{equation}
\dd \oint_{c(\dd g g^{-1})} \hspace{-8mm}\mu/D 
= 
\oint_{c(\dd g g^{-1})} \frac1D \frac{\delta \ell}{\delta a}\diamond a\,dt
\,,
\label{Kel-thm}
\end{equation}
for integration around the material circulation loop $c(\dd g g^{-1})$ moving with stochastic transport velocity $\dd g g^{-1}$. Equation \eqref{Kel-thm} is the Kelvin-Noether theorem from \cite{HoMaRa1998}. As expected, for $\mu/D$ defined in \eqref{v-def}, the Kelvin-Noether circulation theorem now has noise in both its integrand and in its material loop velocity, cf. equations \eqref{refvel-def} and \eqref{VF-1}, respectively. The noise in the integrand is the OU process in \eqref{refvel-def} representing the reference frame velocity, and the stochastic transport velocity of the material loop contains the cylindrical noise in \eqref{VF-1}. 

\begin{remark}[More general forces] \rm More general forces than those in equation \eqref{Kel-thm} may be included into the Kelvin circulation, by deriving it from Newton's Law, as done in \cite{CrFlHo2017}.
\end{remark}

\begin{example}[Euler's fluid equation in 3D]\label{stoch-vort-eqn} \rm 
For an ideal incompressible Euler fluid flow, we have $D=1$ and $\ell(u)=\frac12\|u\|^2_{L^2}$; so $\frac{\delta \ell}{\delta u}=u$  and $D^{-1}\frac{\delta \ell}{\delta a}\diamond a=-\,dp$, where $p$ is the pressure. Upon taking the curl of equation \eqref{mu/D-eqn} in this case and defining total vorticity as $\varpi:={\rm curl}(u+R(x,t))$ we recover the stochastic total vorticity equation,
\begin{equation}
\dd \varpi = {\rm curl} \left((\dd g g^{-1})\times \varpi \right)
= -\, (\dd g g^{-1})\cdot\nabla \varpi + \varpi\cdot\nabla (\dd g g^{-1})
=: - \big[\dd g g^{-1}\,,\,\varpi\big]
=: - \pounds_{\dd g g^{-1}}\varpi
\,,
\label{varpi-eqn-1}
\end{equation}
where
\begin{equation}
\varpi={\rm curl}(u+R(x,t))
\quad\hbox{and}\quad
\dd gg^{-1} = u\,dt + \xi(x)\circ dW_t
\,.
\label{varpi-eqn-2}
\end{equation}
In the latter part of equation \eqref{varpi-eqn-1} we have introduced standard notation for the Lie bracket of vector fields, and identified it as the Lie derivative of one vector field by another.
Thus, the total vorticity in this twice stochastic version of Euler's fluid equation in 3D evolves by the same stochastic Lie transport velocity as introduced in \cite{Holm2015} and analysed in \cite{CrFlHo2017}, but now the total vorticity also has an OU stochastic part, which arises from the curl of the OU stochastic velocity of the large-scale reference frame, relative to which the motion takes place. 

\end{example}

\begin{remark}[It\^o representation] \rm  Spelling out the It\^o representation of these equations will finish Richardson's metaphor, ``And so on, to viscosity'', by introducing the double Lie derivative, or Lie Laplacian, which arises upon writing equation \eqref{mu/D-eqn} in its It\^o form, as in \cite{Holm2015,CrFlHo2017,CrHoRa2016}. For example, the stochastic Euler vorticity equation \eqref{varpi-eqn-1} above is stated
in Stratonovich form. The corresponding It\^{o} form of \eqref{varpi-eqn-1} is
\begin{equation}
\dd\varpi +\pounds_{u}\varpi dt + \mathcal{L}_{\xi}\,\varpi \,dW_{t}
=\frac{1}{2}\pounds_{\xi} ^{2}\varpi\ dt
\,, \label{eq Euler Ito}
\end{equation}
where we write
\begin{equation}
\pounds_{\xi}^{2}\varpi = \pounds_{\xi}(\pounds_{\xi}\varpi)
=\left[  \xi\,,[\xi\,,\,\varpi]\right]  \,,
\label{doubleLieDer}
\end{equation}
for the double Lie bracket of the divergence-free vector field $\xi$ with
the total vorticity vector field $\varpi$. The analysis of the solutions of this 3D vorticity equation in the absence of $R(x,t)$ is carried out in \cite{CrFlHo2017}.  There is much more to report about the dissipative properties of the double Lie derivative in the It\^o form of stochastic fluid equations such as \eqref{eq Euler Ito}. Future work will investigate the combined effects of the OU noise and its interaction with the transport noise  and with the double Lie derivative dissipation in the stochastic total vorticity equation \eqref{eq Euler Ito}.

We will now forego further discussion of the It\^o form of the stochastic Euler fluid equations, in order to continue following the geometric mechanics framework for deterministic continuum dynamics laid out in \cite{HoMaRa1998}, in using the Stratonovich representation to derive the Hamiltonian formulation of dynamics of the stochastic Richardson triple. 

\end{remark}

\begin{remark}[Next steps, other examples, Hamiltonian structure] \rm 
Having established their Hamilton--Pontryagin formulation, the equations of stochastic fluids \eqref{mu/D-eqn} and advection equations \eqref{advect-2} now fit into the Euler--Poincar\'e mathematical framework laid out for deterministic continuum dynamics in \cite{HoMaRa1998}. Further steps in that framework will follow the patterns laid out in \cite{HoMaRa1998} with minor adjustments to incorporate these two types of stochasticity into any fluid theory of interest. In particular, as we shall see, the Lie--Poisson Hamiltonian structure of ideal fluid mechanics is preserved by the introduction of either, or both of the types of stochasticity treated here. 
\end{remark}

\section{Hamiltonian formulation of the stochastic Richardson triple}\label{Ham-form-sec}

\subsection{Legendre transformation to the stochastic Hamiltonian}
The stochastic Hamiltonian is defined by the Legendre transformation of the stochastic reduced Lagrangian in the action integral \eqref{ActionInt-1},  as follows,
\begin{align}
\begin{split}
h(\mu, a)dt &= \Big\langle \mu, \dd gg^{-1} \Big\rangle 
- 
\ell(u,a,D)dt 
- \Big\langle\,DR(x,t)\,,\,u\,\Big\rangle dt
- \Big\langle\,\mu\,,\,\dd gg^{-1} - u\,dt - \xi(x)\circ dW_t\,\Big\rangle
\\&=
\big\langle\,\mu\,,\,u\,dt + \xi(x)\circ dW_t\,\big\rangle
- 
\ell(u,a,D)dt - \big\langle\,DR(x,t)\,,\,u\,\big\rangle dt
\,.
\end{split}
\label{LegXform-1}
\end{align}
Its variations are given by
\begin{align}
\begin{split}
\delta u:\quad&
\frac{\delta h}{\delta u} = \mu - \frac{\delta \ell}{\delta u} - DR(x,t) = 0  \,,
\\
\delta\mu:\quad&
\frac{\delta h}{\delta \mu}dt = u\,dt + \xi(x)\circ dW_t
\,,
\\
\delta a:\quad&
\frac{\delta h}{\delta a}
= - \,\frac{\delta \ell}{\delta a}\,,
\\
\delta D:\quad&
\frac{\delta h}{\delta D}
= - \,\frac{\delta \ell}{\delta D} - \big\langle\,R(x,t)\,,\,u\,\big\rangle
\,.
\end{split}
\label{Ham-var-1}
\end{align}
Consequently, the motion equation in the last line of \eqref{results-var-2} and the advection equations in \eqref{advect-2} may be rewritten equivalently in matrix operator form as
\begin{align}
\begin{bmatrix}
\dd \mu \\ \\ \dd a \\ \\ \dd D
\end{bmatrix}
= -
\begin{bmatrix}
{\rm ad}^*_{(\,\cdot\,)} \mu & (\,\cdot\,)\diamond a & D \nabla (\,\cdot\,)
\\ \\
\pounds_{(\,\cdot\,)}a   & 0 & 0
\\ \\
{\rm div}\big(D(\,\cdot\,)\big) & 0 & 0
\end{bmatrix}\begin{bmatrix}
\frac{\delta h}{\delta \mu}dt \\ \\ \frac{\delta h}{\delta a}dt
\\ \\ \frac{\delta h}{\delta D}dt
\end{bmatrix}.
\label{LPmot-advect-eqns}
\end{align}
The motion and advection equations in \eqref{LPmot-advect-eqns} may also be obtained from the Lie-Poisson bracket for functionals $f$ and $h$ of the flow variables $(\mu,a)$
\begin{equation}
\big\{f,h \big\}
:= 
\left\langle 
\mu\,,\, \left[\frac{\delta f}{\delta \mu}\,,\,\frac{\delta h}{\delta \mu}\right] 
\right\rangle
-
\left\langle 
a\,,\, \pounds_{\frac{\delta f}{\delta \mu}}\frac{\delta h}{\delta a}
-
\pounds_{\frac{\delta h}{\delta \mu}}\frac{\delta f}{\delta a}
\right\rangle
-
\left\langle 
D\,,\, \frac{\delta f}{\delta \mu}\cdot\nabla\frac{\delta h}{\delta D}
-
\frac{\delta h}{\delta \mu}\cdot\nabla\frac{\delta f}{\delta D}
\right\rangle
\,.
\label{LPB-1}
\end{equation}
This is the standard semidirect-product Lie--Poisson bracket for ideal fluids, \cite{HoMaRa1998}. 

Thus, perhaps as expected, the presence of the types of stochasticity introduced into the transport velocity and reference-frame velocity in \eqref{LegXform-1} preserves the Hamiltonian structure of ideal fluid dynamics. However, the Hamiltonian itself becomes stochastic, as in equation \eqref{LegXform-1}, which was obtained from the Legendre transform of the stochastic reduced Lagrangian in the Hamilton--Pontryagin action integral \eqref{ActionInt-1}. This preservation of the Lie--Poisson structure under the introduction of stochastic Lie transport means, for example, that the Casimirs $C(\mu,a)$ of the Lie-Poisson bracket \eqref{LPB-1}, which satisfy $\{C,f\}=0$ for every functional $f(\mu,a)$ \cite{HoMaRa1998}, will still be conserved in the presence of stochasticity. In turn, this conservation of the Casimirs implies the preservation of the standard potential vorticity (PV) invariants, even for stochastic Lie transport dynamics of fluids in a randomly moving reference frame, as treated here. However, as mentioned earlier, the total energy and momentum will not be preserved, in general, because their correspond Noether symmetries of time and space translation have been violated in introducing explicit space and time dependence in the stochastic parameterisations of the transport velocity and reference frame velocity. 

\subsection{Gyroscopic analogy}
There is a well-known gyroscopic analogy between the spatial moment equations for stratified Boussinesq fluid dynamics and the equations of motion of the classical heavy top, as discussed e.g., in \cite{Holm1986,Do2013} and references therein. This analogy exists because the dynamics of both systems are based on the same semidirect-product Lie-Poisson Hamiltonian structure. One difference between the spatial moment equations for Boussinesq fluids and the equations for the motion of the classical heavy top is that for fluids 
the spatial angular velocity is right-invariant under $SO(3)$ rotations, while the body angular velocity for the heavy top is left-invariant under $SO(3)$ rotations.  Because the conventions for the heavy top are more familiar to most readers than those for the Boussinesq fluid gyroscopic analogy, and to illustrate, in passing, the differences between left-invariance and right-invariance, we shall discuss the introduction of OU reference frame stochasticity for the classical heavy top in this example.%
\footnote{The introduction of stochastic Lie transport for the classical heavy top has already been investigated in \cite{ArCaHo2017}.}

\begin{example}[Heavy top] \rm 
The action integral for a heavy top corresponding to the fluid case in \eqref{ActionInt-1} is given by,
\begin{equation}
S = \int \ell\big(\Omega(t),O^{-1}(t)\boldsymbol{\widehat{z}}\big)dt + \langle\,R(t)\,,\,\Omega(t)\,\rangle dt
+
\langle\,\Pi\,,\,O^{-1}\dd O(t) - \Omega\,dt - \xi\circ dW_t\,\rangle\,.
\label{ActionInt-2}
\end{equation}
with $\Omega(t)\in\mathfrak{so}(3)\simeq\mathbb{R}^3$, $O(t)$ a curve in $SO(3)$, $R(t) = \eta N(t)\in \mathbb{R}^3$, $\boldsymbol{\widehat{z}}=(0,0,1)^T$ the vertical unit vector, $\Pi(t)\in\mathfrak{so}(3)^*\simeq\mathbb{R}^3$, constant vectors $\eta,\xi\in \mathbb{R}^3$, and the brackets $\langle \,\cdot\,,\,\cdot\,\rangle$ denote $\mathbb{R}^3$ pairing of vectors. After the Legendre transform, the Hamiltonian is determined as in \eqref{LegXform-1} to be
\begin{align}
h(\Pi, \Gamma)dt =
\big\langle\,\Pi\,,\,\Omega\,dt + \xi\circ dW_t\,\big\rangle
- 
\ell(\Omega,\Gamma)dt - \big\langle\,R(t)\,,\,\Omega\,\big\rangle dt
\,,
\label{LegXform-2}
\end{align}
with $\Gamma(t):=O^{-1}(t)\boldsymbol{\widehat{z}}$. The variations are given by
\begin{align}
\begin{split}
\delta \Omega:\quad&
\frac{\delta h}{\delta \Omega} = \Pi - \frac{\delta \ell}{\delta \Omega} - R(t) = 0  \,,
\\
\delta\Pi:\quad&
\frac{\delta h}{\delta \Pi}dt = \Omega\,dt + \xi\circ dW_t
\,,
\\
\delta \Gamma:\quad&
\frac{\delta h}{\delta \Gamma}
= - \,\frac{\delta \ell}{\delta \Gamma}
\,.
\end{split}
\label{Ham-var-2}
\end{align}
The motion equation for $\Pi$ and the advection equation for $\Gamma$ may be rewritten equivalently in matrix operator form as
\begin{align}
\begin{bmatrix}
\dd \Pi \\ \\ \dd \Gamma
\end{bmatrix}
= 
\begin{bmatrix}
\Pi \times  & \Gamma \times 
\\ \\
\Gamma \times    & 0
\end{bmatrix}
\begin{bmatrix}
\frac{\partial h}{\partial \Pi}dt 
\\ \\ 
\frac{\partial h}{\partial \Gamma}dt
\end{bmatrix}
=
\begin{bmatrix}
\Pi \times \frac{\partial h}{\partial \Pi}dt  
+  \Gamma \times \frac{\partial h}{\partial \Gamma}dt
\\ \\ 
\Gamma \times \frac{\partial h}{\partial \Pi}dt  
\end{bmatrix}
\label{LPmot-HT-eqns}
\end{align}
The motion and advection equations in \eqref{LPmot-HT-eqns} may also be obtained from the Lie-Poisson bracket for functionals $f$ and $h$ of the  variables $(\Pi,a)$
\begin{equation}
\big\{f,h \big\}
:= 
-\left\langle 
\Pi\,,\, \frac{\delta f}{\delta \Pi}\times\frac{\delta h}{\delta \Pi}
\right\rangle
-
\left\langle 
\Gamma\,,\, \frac{\delta f}{\delta \Pi}\times \frac{\delta h}{\delta \Gamma}
-
\frac{\delta h}{\delta \Pi}\times\frac{\delta f}{\delta \Gamma}
\right\rangle
\,, 
\label{LPB-2}
\end{equation}
which is the standard semidirect-product Lie--Poisson bracket for the heavy top, with $(\Pi,\Gamma)\in\mathfrak{se}(3)^*\simeq (\mathfrak{so}(3)\circledS\mathbb{R}^3)^*$, where $\circledS$ denotes semidirect product. 

The deterministic Lagrangian for the heavy top is 
\begin{align}
\ell(\Omega,\Gamma) = \frac12 \Omega \cdot I\Omega 
- mg\chi\cdot \Gamma
\,,
\label{Lag-3}
\end{align}
where $I$ is the moment of inertia, $m$ is mass, $g$ is gravity and $\chi$ is a vector fixed in the body.
The corresponding stochastic Hamiltonian, obtained by specialising the Hamiltonian in \eqref{LegXform-2} is \cite{Holm2011},
\begin{align}
h(\Pi, \Gamma)dt =
\big\langle\,\Pi\,,\,\Omega\,dt + \xi\circ dW_t\,\big\rangle
- \frac12 \Omega \cdot I\Omega\,dt + mg\chi\cdot \Gamma dt
 - \big\langle\,R(t)\,,\,\Omega\,\big\rangle dt
\,,
\label{LegXform-3}
\end{align}
whose variations are given by
\begin{align}
\begin{split}
\delta \Omega:\quad&
\frac{\delta h}{\delta \Omega} = \Pi - I\Omega - R(t) = 0  \,,
\\
\delta\Pi:\quad&
\frac{\delta h}{\delta \Pi}dt 
= \Omega\,dt + \xi\circ dW_t
= I^{-1}\big(\Pi - R(t)\big)dt + \xi\circ dW_t 
\,,
\\
\delta \Gamma:\quad&
\frac{\delta h}{\delta \Gamma}
= - \,mg\chi
\,.
\end{split}
\label{Ham-var-2}
\end{align}
Thus, according to equations \eqref{LPmot-HT-eqns} we have, with $R(t) = \eta N(t)\in \mathbb{R}^3$, 
\begin{align}
\begin{bmatrix}
\dd \Pi \\ \\ \dd \Gamma
\end{bmatrix}
= 
\begin{bmatrix}
\Pi \times  & \Gamma \times 
\\ \\
\Gamma \times    & 0
\end{bmatrix}
\begin{bmatrix}
I^{-1}\Pi\,dt - I^{-1}R(t)dt + \xi\circ dW_t 
\\ \\ 
- \,mg\chi dt
\end{bmatrix}.
\label{LPmotHT-eqns-1}
\end{align}
Thus, in the dynamics of angular momentum, $\Pi$, the effects of two types of noise add together in the motion equations for the Stratonovich stochastic heavy top in an OU rotating frame. 

In terms of angular velocity, $\Omega$, equation \eqref{LPmotHT-eqns-1} may be compared more easily with its fluid counterpart in \eqref{mu/D-eqn},
\begin{align}
\begin{bmatrix}
\dd\big(I \Omega + R(t) \big) \\ \\ \dd \Gamma
\end{bmatrix}
= 
\begin{bmatrix}
I\dd \Omega + \eta\big[\theta (\overline{N} - N(t))\,dt + \sigma dW_t \big] \\ \\ \dd \Gamma
\end{bmatrix}
= 
\begin{bmatrix}
\big(I\Omega + R(t) \big)\times  & \Gamma \times 
\\ \\
\Gamma \times    & 0
\end{bmatrix}
\begin{bmatrix}
\Omega\,dt +  \xi dW_t 
\\ \\ 
- \,mg\chi dt
\end{bmatrix},
\label{LPmotHT-eqns-2}
\end{align}
upon substituting  $\dd N = \theta (\overline{N} - N(t))\,dt + \sigma dW_t$ from \eqref{OU-1}. In equation \eqref{LPmotHT-eqns-2}, one sees that the deterministic angular momentum $I\Omega$ has been enhanced by adding the integrated OU process $R=\eta N(t)$ with $N(t)$ given in \eqref{OUsoln-1}, while the deterministic angular transport velocity $\Omega$ is enhanced by adding the Stratonovich noise. However, one need not distinguish between It\^o and Stratonovich noise in \eqref{LPmotHT-eqns-2}, since $\eta$ and $\xi$ are both constant vectors in $\mathbb{R}^3$ for the heavy top. The stochastic equation \eqref{LPmotHT-eqns-2} in the absence of the OU noise, $R(t)$, was studied in \cite{ArCaHo2017} to investigate the effects of transport stochasticity in the heavy top. Future work reported elsewhere will investigate the combined effects of the OU noise and its interaction with the transport noise in the examples of the rigid body and heavy top. 

\end{example}

\section{Conclusion/Summary}

In this paper we have defined the Richardson triple as an ideal fluid flow map $g_{t/\ep,t,\ep t} = h_{t/\ep}k_t l_{\ep t}$ composed of three smooth maps with separated time scales. We then recast this flow as a single SPDE and expressed its solution behaviour in terms of the Kelvin circulation theorem \eqref{Kel-thm} involving two different stochastic parameterisations. 

The first type of stochastic parameterisation we have considered here appears in the Kelvin circulation \emph{loop} in \eqref{Kel-thm}, as ``Brownian transport''  of the resolved intermediate scales by the unresolved smaller scales. The stochastic model for this type of transport was introduced in \cite{Holm2015}, and was later derived constructively by employing homogenisation in \cite{CoGoHo2017}. Its solution properties for the 3D Euler fluid were analysed mathematically in \cite{CrFlHo2017} and its transport properties were developed further for applications in geophysical fluid dynamics involving time-dependent correlation statistics in \cite{GBHo2017}.  The inference of the stochastic model parameters in two applications of the class of Stochastic Advection by Lie Transport (SALT) fluid models discussed here has been implemented recently. The implementations for proof of principle of this type of stochastic parameterisation in numerical simulations have been accomplished for Euler's equation for 2D incompressible fluid flows in \cite{Cotter-etal-2018a} and for simulations in 2D of 2-layer quasigeostrophic equations in \cite{Cotter-etal-2018b}.

The second type of stochastic parameterisation we have considered here appears in the Kelvin circulation \emph{integrand} in \eqref{Kel-thm}, and it accounts for the otherwise unknown effects of stochastic non-inertial forces on the stochastic dynamics of the composite intermediate and smaller scales, relative to the stochastically moving reference frame of the larger whorl. Of course, the physics of moving reference frames is central in deterministic GFD and it goes back to Coriolis for particle dynamics in a rotating frame. However, judging from the striking results of \cite{CCEH2005} and \cite{GeHoKu2007} for transport in rotating turbulence, we expect that the introduction of non-inertial stochasticity could have interesting physical effects; for example, on mixing of fluids, especially because the magnitude of the rotation rate, for example, is not limited in the SALT formulation. 

Section \ref{Var-form-sec} established the variational formulation of the SPDE representation of the stochastic Richardson triple, thereby placing it into the modern mathematical context of geometric mechanics \cite{Arnold1966,HoMaRa1998}. 
Having established their Hamilton--Pontryagin formulation, the equations of stochastic fluids \eqref{mu/D-eqn} and advection equations \eqref{advect-2} now fit into the Euler--Poincar\'e mathematical framework laid out for deterministic continuum dynamics in \cite{HoMaRa1998}. In this framework, the geometrical ideas of \cite{Arnold1966}, in which ideal Euler incompressible fluid dynamics is recognised as geodesic motion on the Lie group of volume preserving diffeomorphisms may be augmented to include advected fluid quantities, such as heat and mass. The presence of advected fluid quantities breaks the symmetry of the Lagrangian in Hamilton's principle under the volume preserving diffeos in two ways, one that enlarges  the symmetry, and another that reduces the symmetry. First, compressibility enlarges the symmetry group to the full group of diffeos, to include evolution of fluid volume elements. Second, the initial conditions for the flows with advected quantities reduce the symmetry under the full group of diffeos to the subgroups of diffeos which preserve the initial configurations of the advected quantities. These are the \emph{isotropy subgroups} of the diffeos. Boundary conditions also reduce the symmetry to a subgroup, but that is already known in the case of \cite{Arnold1966} for the incompressible flows. In the deterministic case, the breaking of a symmetry to an isotropy subgroup of a physical order parameter leads in general to semidirect-product Lie--Poisson Hamiltonian structure \cite{Holm2011,HoMaRa1998}. This kind of Hamiltonian structure is the signature of symmetry breaking. One might inquire whether the semidirect-product Hamiltonian structure for deterministic fluid dynamics would be preserved when stochasticity is introduced. 

Section \ref{Ham-form-sec} revealed that this semidirect-product structure on the Hamiltonian side is indeed preserved for the two types of stochastic deformations we have introduced here. In addition, we saw that either of these two types of stochastic deformation may be introduced independently. Section \ref{Ham-form-sec}  also provided an example of the application of these two types of stochasticity in the familiar case of the heavy top, which is the classical example in finite dimensions of how symmetry breaking in geometric mechanics leads to semidirect-product Lie--Poisson Hamiltonian structure. For geophysical fluid dynamics (GFD) the preservation of the semidirect-product Lie--Poisson structure under the introduction of the two types of stochasticity treated here implies the preservation of the standard potential vorticity (PV) invariants for stochastic Lie-transport dynamics of fluids, suitably modified to account for non-inertail forces for motion relative to a randomly moving reference frame. 

As mentioned in the introduction, we intend that the approach described here in the context of the Richardson triple will be useful for the stochastic parameterisation of uncertain multiple time scale effects in GFD. For example, it may be useful in quantifying predictability and variability of sub-mesoscale ocean flow dynamics in the random frame of one or several unspecified mesoscale eddies interacting with each other. We also hope this approach could be useful in investigating industrial flows, e.g., to quantify uncertainty and variability in multi-time-scale flow processes employed in industry.

\subsubsection*{Acknowledgements.} The author is grateful for stimulating and thoughtful discussions during the course of this work with D. O. Crisan, C. J. Cotter, F. Flandoli, B. J. Geurts, R. Ivanov, E. Luesink, E. M\'emin, W. Pan, V. Ressiguier and C. Tronci. The author is also grateful for partial support by the EPSRC Standard Grant EP/N023781/1.

\end{document}